\documentclass{article}

\usepackage{amsmath,amsthm,amssymb}
\usepackage{url}
\usepackage{subfigure}
\usepackage{graphicx}
\usepackage{amsmath,amssymb}
\usepackage{amstext}
\usepackage{dsfont}
\usepackage{xspace}
\usepackage{epsfig}
\usepackage{balance}
\usepackage{color}
\definecolor{lightgray}{cmyk}{0.0,0.0,0.0,0.06}
\definecolor{darkgray}{rgb}{0.36,0.36,0.36}

\usepackage[plainpages=false,pdfpagelabels]{hyperref}

\begin{document}

\title{Dynamic Integration of ALM Tools for\\Agile Software Development}

\author{Max Wielsch, Raik Bieniek, Bernd Grams\\
Saxonia Systems AG\\
D-01069 Dresden, Germany
\and J{\"o}rg L{\"a}ssig\\
University of Applied Sciences Zittau/G\"orlitz\\
D-02826 G\"orlitz, Germany
}

\maketitle
\begin{abstract}
The paper describes the need for and goals of tool-integration within software development processes. In particular we focus on agile software development but are not limited to. The integration of tools and data between the different domains of the process is essential for an efficient, effective and customized software development. We describe what the next steps in the pursuit of integration are and how major goals can be achieved. Beyond theoretical and architectural considerations we describe the prototypical implementation of an open platform approach. The paper introduces platform apps and a functionality store as general concepts to make apps and their functionalities available to the community. We describe the implementation of the approach and how it can be practically utilized. The description is based on one major use case and further steps are motivated by various other examples.
\end{abstract}

\section{INTRODUCTION}\label{sec:introduction}

Software development is a complex and expensive procedure. 
Nowadays, there is an ample number of different specialized tools available to support the software development process --or various kinds thereof-- best possible. In the recent years, one could notice significant changes in the capabilities and features of those tools - a continuous development towards more integrated tool suites and platforms took place. Tools are integrating better and closer with each other and thereby allow to connect more and more formerly separated parts of the process chains together in an interoperating manner.

Only years ago, more static process models like the Waterfall model or the V-Model have been used heavily as a standard approach. Today, agile models and methods like Scrum or Kanban are spreading in usage and are currently state of the art. Triggered by this change of the applied process models, continual improvement of the software development as major inherent characteristics becomes more and more evident. Appreciably, vendors do spend much efforts to improve their tools to support this development better. Further, communities like OSLC\footnote{Open Services for Lifecycle Collaboration,~\url{http://open-services.net}.} focus on creating  standards for sharing and exchanging data between different domains~\cite{berger2011,zetie2009}. As a result, shorter release cycles and the fulfillment of rising demands on quality or goals as an improved interplay between development and IT operations can be targeted.

Today, tools are able to connect lifecycle artifacts like requirements and test cases. E.\,g., modern developers are provided with functionalities to trace changes made in different development domains. There are tools for quality management (Quality Center, Tosca), continuous integration (Jenkins, TFS, Bamboo), software configuration management and version control (Git, TFS, ClearCase) and other domains. Many of them are assembled together in tool suites such as the Atlassian Suite, IBM Rational Suite, Microsoft TFS or integrated in platforms like Tasktop Sync or Polarion ALM.

In order to capture upcoming steps of development and our vision thereof in the continuous improvement process of tool support, we focus subsequently on the stage where a new software project is started, i.\,e., on the setup phase of a new project. Every project comes with its own demands and requirements depending on the customer, the companies’ goals, the developers’ needs or even on laws and policies. Regarding the special circumstances, the project including the development environment should be set up differently under  different circumstances. For each demand, a special tool is chosen and configured. Tools may be offered by third-party vendors or developed by the developer team or developing company itself. Given this level of supporting tools and integration, the project initiation procedure can be executed quickly, as there are pre-configured tools that only have to be adjusted to the special needs. As the development is done in an agile way, the development team has to be able to respond to new situations and new requirements quickly. 

Given this environment, there is no need to deal with workarounds or to develop  inefficiently. Also, not only the development team obtains the needed functionality or (lifecycle) data from sub-processes and its repositories given this environment; also the quality assurance team or project managers get instantly the resources needed in order to work efficiently. The project manager obtains updated information instantaneously and statistics provide him with the ability to make decisions precise and timely. The Test Manager recognizes test cases to be changed immediately after a requirement has been changed, as all lifecycle data is connected and workflows as requirement changes are automated.

At least partly, the described environment is still a vision so far but it can be realized by advancing integration technologies~\cite{eskeli2010}. This work describes an approach to achieve this by designing and implementing an open development integration platform --\emph{OpenDIP}-- which aims to achieve the vision. % described above. 

The rest of the paper is structured as follows. In the next section, we further motivate our work with typical useful applications for the software development process, that can be realized by applying the open platform described in this paper. Then, in Section~\ref{sec:relatedWork}, previous work on the topic is reviewed. Section~\ref{sec:OpenddipPlatformConception} infers the requirements of the platform based on an exemplary use case - the Timeline App. Then, in Section~\ref{sec:architecture}, beginning with an architectural overview the main components of the platform are described. The platform meta model as a core feature of the platform is introduced. With the knowledge of basic platform concepts and the architecture thereof it is discussed what steps have to be done to realize an app in Section\ref{sec:DevelopingPlatformApps}. The description is again based on the Timeline App. It serves as simple practical example for a platform app, as representative of the platform apps which are described in Section~\ref{sec:Motivation}. Section~\ref{sec:Conclusions} concludes the paper and in Section~\ref{sec:FutureWork} the next steps in our project are motivated.

%%%%%%%%%%%%%%%%%%%%%%%%%%%%%%%%%%%%%%%%%%%%%%%%%%%%%%%%%%%%%%%%%%%%%%%%%%%%%%%%

\section{Motivation}\label{sec:Motivation}

The following examples of applications are described in short to demonstrate the spectrum of possibilities that may be derived from the properties of the integration platform which is described in the forthcoming paragraphs -- both by tool manufacturers and development teams.

\paragraph{Information Dashboard} The dashboard app provides information about the development process for the different stakeholder roles in real time. Key figures like performance and quality indicators are calculated from the custom data model hold within the platform. The data is presented in living graphics, i.\,e. the diagrams are refreshed by update events from the data model and do not have to be newly created periodically to refresh data.

\paragraph{Project Database} The project database app is used to automatically collect data from projects and user stories over time. This data storage is a valuable treasure that can be used by the team to improve the estimation of efforts or to reproduce prior solutions to implementation problems.

\paragraph{Project Cockpit} The cockpit app is a central access point with single sign-on to the different tools in use. Here the developer can start project builds or run tests and watch the build status of projects and the results of test executions without the need to switch between tools.

\paragraph{Quality Assurance Workflows} These workflows are scripts developed by the team using a dynamic language like JavaScript to automate processes. These workflows that run directly in the platform environment can be used to record user or system activities, store documents and send emails triggered by certain events to meet the requirements of ISO 9001 and other quality standards. Domains with special demands for the software development processes as for instance manufacturers of medical instruments can develop and run their own automated processes. Both, the workflow scripts and the data models, can be exchanged between different platforms in use.

\begin{figure}[h!]
  \color{darkgray}\fbox{\includegraphics[scale=0.4]{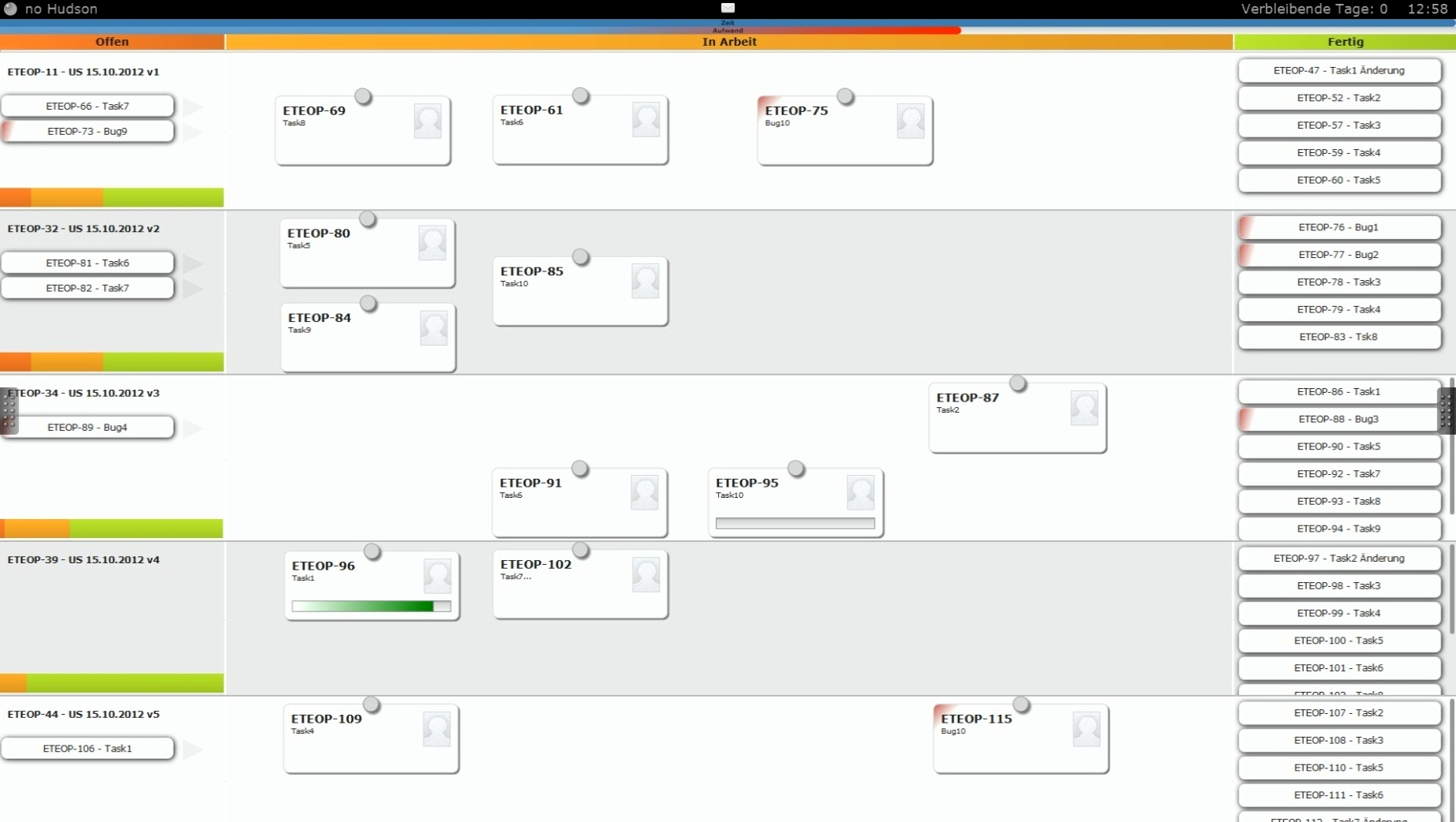}}\color{black}
  \centering
  \caption{Prototype of the Digital Scrumboard. Typically a Scrumboard consists of the three areas \emph{To Do, In Progress} and \emph{Done}, signaling the status of tasks of a sprint. The tasks are synchronized with an issue tracking system.}
  \label{fig:scrumboard}
\end{figure}

\paragraph{Digital Scrumboard} The digital scrumboard is a graphical front-end to an issue tracking system that is used to manage the user stories and tasks to be implemented by a developer team. The status of a task on the board is updated by events from the issue tracking tool. Vice versa, the manual change of the status of a task on the board is automatically stored to the issue tracking system. With the digital scrumboard app the team can use a huge touch screen for the daily scrum meetings with a lot of advantages over a conventional pinboard. By connecting the data models across virtual machines, multiple scrumboards can be synchronized to support development in distributed teams, see Figure~\ref{fig:scrumboard}.   

\setlength{\fboxsep}{0.0pt}
\setlength\fboxrule{0.75pt}

\paragraph{Consistency Checker} This app checks the consistency of data across tool borders using rules describing functional conditions for consistent data. It is possible to formulate queries like "Show me all requirements without any test case." or "Which test cases have to be adjusted 
when requirement xyz will be deleted?".

\section{RELATED WORK}
\label{sec:relatedWork}

Integrating applications, in particular developments tools, is a topic that has been investigated already for a long time~\cite{brown1992,thomas1992}. Thomas~\cite{thomas1992} and Brown~\cite{brown1992} give a general and abstract perspective on the topic and introduce very common goals and directions of integration, which are namely: \emph{presentation integration} (the user's mental model should be supported by the presentation of information), \emph{data integration} (related to the aspects interoperability, non-redundancy, data consistency, data exchange and synchronization), \emph{control integration} (related to the provisioning and usage of tool functionality) and \emph{process integration} (the extend to which the tooling supports process properties like steps, events and constraints). The authors point out that those properties are not meant to be fulfilled by a single tool to integrate, but by the relationships of the different tools used. So finally, Thomas introduces a beneficial metric of integration properties for evaluating the integration properties of the framework introduced within this work.

Further, there are investigations describing integration solutions with similar goals as our solution~\cite{sinha2007,frantz2012}. Sinha \emph{et al.}~\cite{sinha2007} elaborate on the issues of today's software development. Additionally to the motivation for integration as emphasized by Brown and Thomas they mention the increasingly distributed development as new reason of the need for integration. They describe a conceptual framework (with an abstract architecture) that partly matches the ideas and concepts mentioned in Section~\ref{sec:introduction}. Using several real-life examples they argue for the need of an integration framework. 

Meanwhile, Frantz and Corchuelo~\cite{frantz2012} refer to integration in the business (EAI) context. But, they state important facts that also apply to integration of the software development process. They see need for integration in order to be able to reuse software components (or tools) and for adapting or optimizing IT to the needs of (business) processes, while we emphasize workflows within the software development process. Further, they introduce their own EAI framework to compare it to other existing ones like Spring Integration or Camel. With regard to the goal to provide simple tool integration and development mechanisms, we consider the approach to be not sufficient to serve as an  integration environment for software development processes. In particular the use of message pattern to detach data from integrated applications can be problematic. This means, messages have to be routed through the integration environment and translated into the application specific data model format. Using message buses to decouple systems is a general EAI pattern, which we explicitly avoid by mapping directly on a generic meta model understood by all applications using our platform.

Another approach to be mentioned here is pursued by Biehl \emph{et al.}~\cite{biehl2012}. They build a service discovery and orchestration framework for OSLC services, representing single development tools and systems. The idea of OSLC is to provide vendor independently lifecycle data -- originally for the application lifecycle management (ALM). For orchestrating tools' functionalities, exposed via RESTful OSLC services, process chains are described by a domain specific language. Decoupling vendor specific models is a very vital feature of OSLC, but it has also its drawbacks due to the use of RESTful services. Mapping bidirectional communication with RESTful services is --although possible with workarounds-- by definition not considered in HTTP.

Nevertheless, neither is integration of applications and lifecycle data a very new problem nor it is restricted to the integration of lifecycle data of applications, exclusively. Analogous to the ALM integration of lifecycle data, the integration of other lifecycle data is treated by other branches, e.\,g. for product lifecycle management (PLM). That is, the integration trend exceeds the domain of software development. Srinivasan~\cite{srinivasan2011} describes for the domain of PLM an integration framework using standardized product data, meta-data models and standardized business processes, on the one hand, and on the other hand a service oriented architecture concept to integrate different tools and systems. 

Arguing for the need of continuing with advancing integration technologies, Seligman \emph{et al.}~\cite{seligman2010} determine that existing tools --although already advanced-- are still too costly and labor intensive to build. Their solution is presented in form of a platform. The platform consists of a (vendor) model-independent repository, containing model schemas and a mapping of integrated tools, data importing and exporting components and integration tools using the eclipse platform and its plug-in mechanism. Considering the architecture of Seligman \emph{et al.}, a platform feature for easy and dynamic usage of integrated functionalities and data is still missing.

To summarize, we conclude many authors consider application integration and in particular the integration of the software development process, pointing out that a general and satisfying solution of this problem is still missing. Not all so far applied concepts can be explicitly mentioned here - we restrict ourselves to a few major characteristics and facts which are important for our platform design.

%%%%%%%%%%%%%%%%%%%%%%%%%%%%%%%%%%%%%%%%%%%%%%%%%%%%%%%%%%%%%%%%%%%%%%%%%%%%%%%%

\section{OPENDIP PLATFORM CONCEPTION}
\label{sec:OpenddipPlatformConception}

With the vision of future directions in software development as described in the introduction and the motivation, we tried to outline some characteristics of the development process that we aim for with OpenDIP. In this section we deepen our considerations with concrete architectural concepts and entities.

\begin{figure}
  \centering
  \includegraphics[scale=0.55]{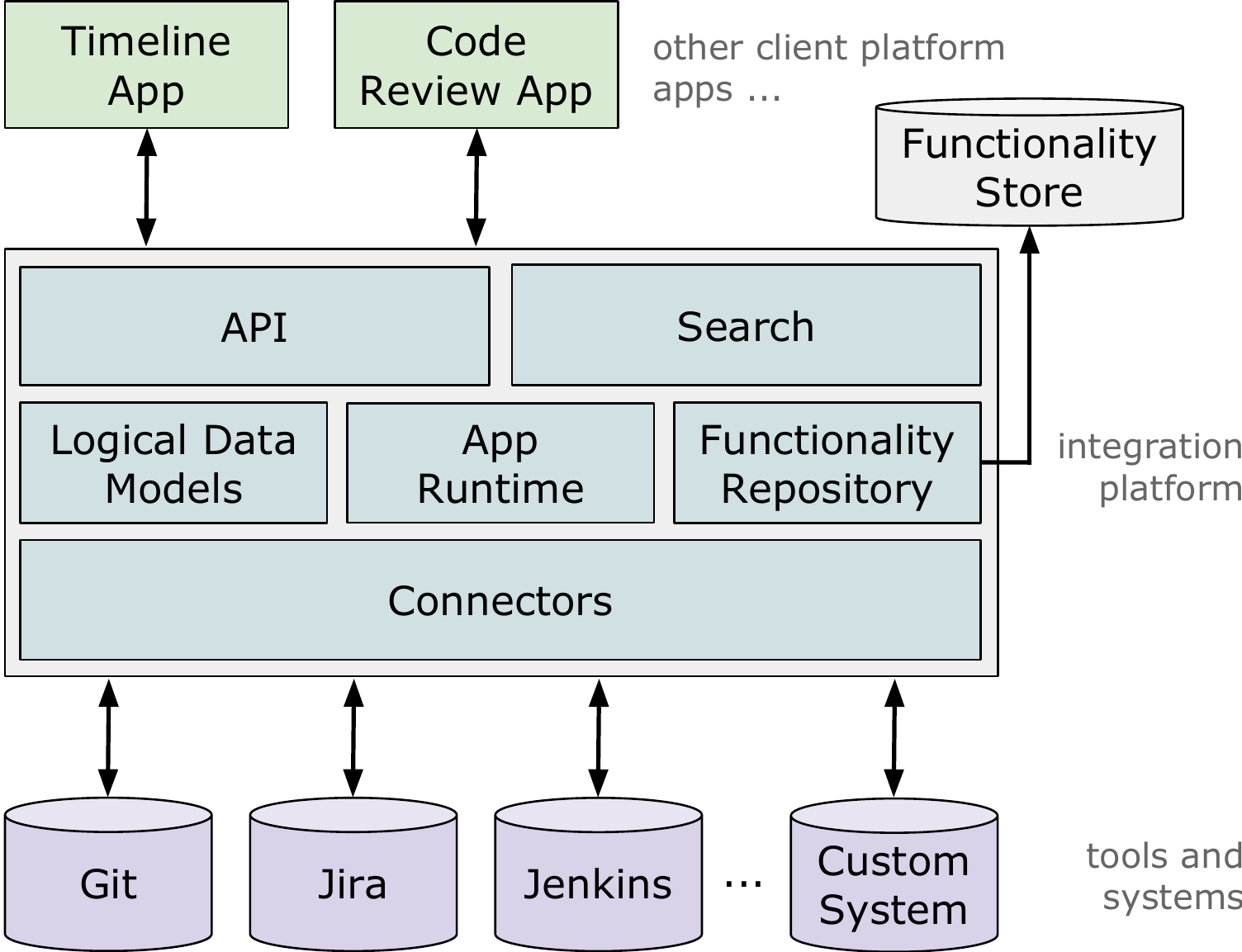}
  \caption{The platform ecosystem is made up of four main parts. Tools and systems offered by third-parties or custom systems (purple), the platform itself (gray) and client platform apps to support the software development process (green). Further, there is a functionality store that is connected to the platform.}
  \label{fig:opendip_concept}
\end{figure}

In the following, we refer to Figure~\ref{fig:opendip_concept} to explain the basic platform requirements and concepts.

Four main entities of interest are visualized. \emph{Tools and systems} (third-party or custom ones) needed in the software development process (purple), the \emph{platform} itself (gray), whose features are depicted inside with blue boxes, and \emph{client platform apps} to use the integrated functionality of the platform for various special purposes (green). Further, there is a \emph{functionality store} that is connected to the platform, containing apps which have been developed by the community for the community.

All connected custom and third-party tools provide resources -- functionality and lifecycle data -- that needs to be integrated.
The range of tools or systems that can be included should not be limited to traditional ALM tooling. There are many more tools, applications or systems providing specific functionalities and data to be used within the development process. E.\,g., there are also communication tools like instant messengers or voice over IP services, web applications or cloud services as Dropbox or Stackoverflow and more. For all of those having an API or a CLI, the platform should provide an integration mechanism, e.\,g. via \emph{connectors}. Communication between those integrated systems or tools, the platform and client tools, should be bidirectional. Many scenarios of real world processes contain bidirectional relationships.
The (client-side) platform apps (on top of the figure) support special or custom business cases.

The \emph{Timeline app} is a representative for a useful client app, leveraging the development process -- further client apps are described in Section~\ref{sec:Motivation}. The utility of this app here is overviewing all events of the process allowing to react on those instantly. Events are provided from several software systems connected to the platform, like systems for source code management, issue management systems, continuous integration servers, and others.

For developing these platform apps, data and functionality of integrated systems have to be provided.
Hence, an \emph{API} for accessing data and functionality uniformly is needed. An easy-to-use extension mechanism is here a crucial distinguishing factor compared to other existing/commercial platforms. The platform must be extensible by custom data models (\emph{logical data models}) and functionality. Hence, vendor specific models are decoupled from platform client apps. For the purpose of collecting, representing and reusing custom functionality and data models, the platform must have an internal \emph{functionality repository}, which is connected to a functionality store. 
The functionality store is an external repository used to provide and distribute already created functionalities -- and apps connected to it.
Functionalities can be used for implementation purposes, apps are ready to use. If some functionality is missing in the internal repository (functionalities already available in the running platform/environment), it might be obtained from the functionality store. The functionality store can be queried by users of the platform (software developers), searching for new functionalities needed in the software development process. In the ideal case, an app providing this functionality is found directly. The community helps to further develop the range and number of available functionalities. Besides client apps that are executed as a separate application, there are also platform apps that will be executed by the platform on the \emph{app runtime} environment. An important use case for the app runtime are automated workflows. Those are executed like other platform apps in the app runtime too. 

A typical use case for systems dealing with available platform data and functionalities is \emph{search}.
The platform provides a search function that not only searches snippets in texts for equality. Also the range of much different model data of connected systems needs to be searchable.
Hence, search should be executed on semantic information.
Consequently, the platform must allow for enriching underlying data models with semantics.
Another important platform characteristics is that users (developers) should be able to create spontaneously needed functionality dynamically, without the need to restart the platform. For implementing such \emph{dynamic functionalities}, which are also treated as an app, a scripting API is the right means. App scripts can be executed within a runtime environment of the platform. 

This covers only major requirements of the platform, there are many more to be considered but because of space restrictions we focus here only on core features.

%%%%%%%%%%%%%%%%%%%%%%%%%%%%%%%%%%%%%%%%%%%%%%%%%%%%%%%%%%%%%%%%%%%%%%%%%%%%%%%%

\section{PLATFORM ARCHITECTURE}\label{sec:architecture}

One of the main requirements of integration platforms in general is to provide mechanisms to share data between connected systems. 
To support such mechanisms, an abstract architecture for data sharing components is developed, which is presented in Section~\ref{platformArchitectureOverview}.
A key concept of that architecture is that shareable data is represented based on an entity meta model which is described in Section~\ref{platformArchitectureMetaModel}.

\subsection{Overview}\label{platformArchitectureOverview}

\begin{figure}[h!]
  \includegraphics[scale=0.585]{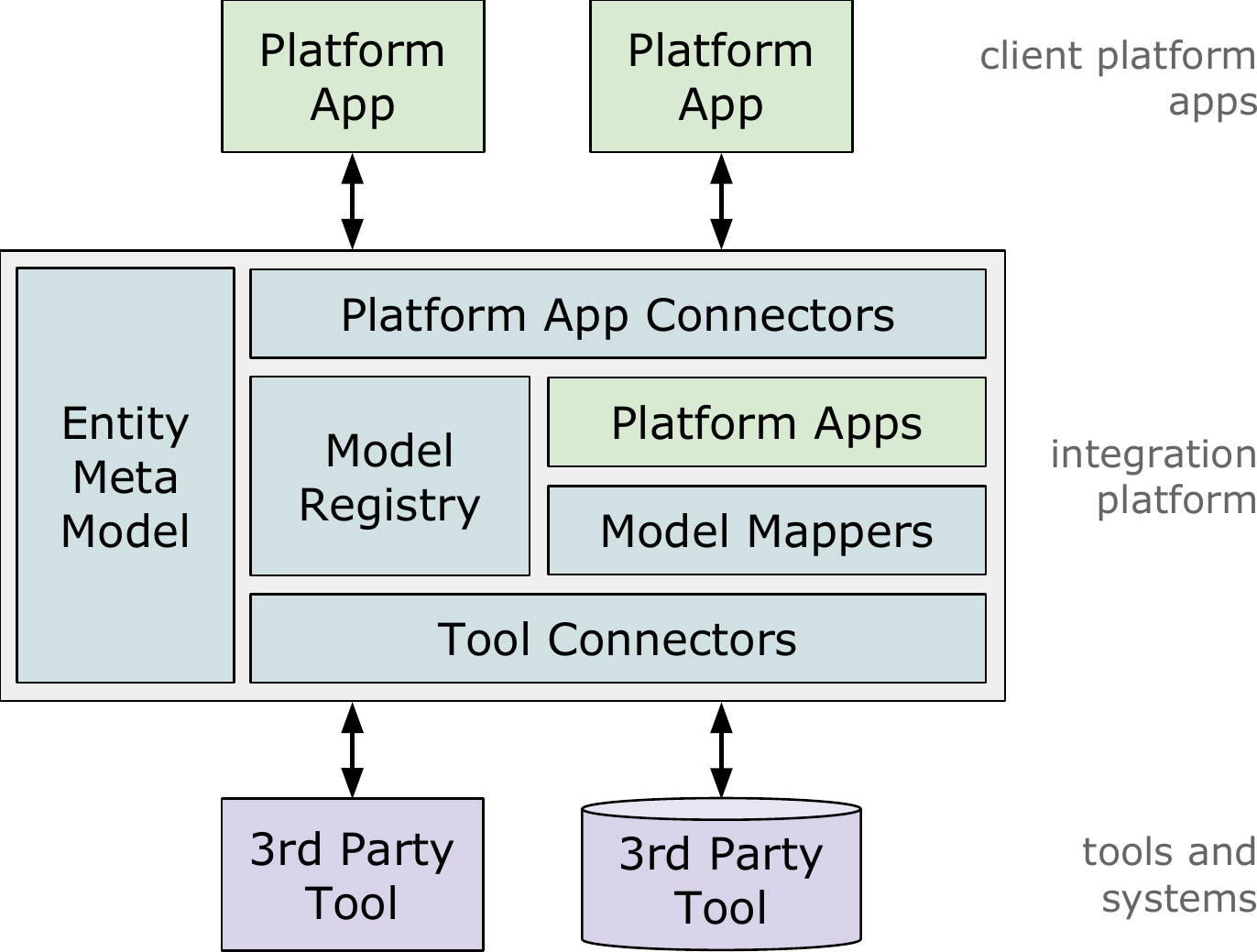}
  \centering
  \caption{The abstract architecture of the platform. Tool connectors link to external (custom or third-party) tools and make their data available via the model registry. Model mappers translate between data models and platform apps which use these data models for their business use cases.}
  \label{fig:abstract_architecture}
\end{figure}

Figure~\ref{fig:abstract_architecture} gives a short overview of the abstract architecture of the platform.
Platform  modules (blue) work with structured data complying with some domain model.
Each module which makes domain model instances available to other modules does so by registering the root element of these instances in a special module called \emph{model registry} (Figure~\ref{fig:abstract_architecture}).
Modules accessing data of a domain model instance must retrieve the root element of the instance from the model registry first to be able to navigate to the desired data from there.

For the purpose of integrating third-party \emph{tools and systems} (Figure~\ref{fig:abstract_architecture}) not adhering to the concept of this data sharing architecture,  tool connectors are used.
In case of the Timeline app, connectors for ALM systems, like an issue management system (e.\,g. Track) or a build server (e.\,g. Jenkins), need to be provided.
More generic systems, like database management systems, are also considered third-party systems.
The task of \emph{tool connectors} (Figure~\ref{fig:abstract_architecture}) is to access data of these third-party tools and to translate it to domain model instances basing on the \emph{entity meta model} (Figure~\ref{fig:abstract_architecture}).
This translation must ensure that every piece of data that is requested from the connector is live retrieved from the integrated third-party system.

Also, the connector must forward data that is written to the translated meta model instance backwards to the third-party tool.
That is, the tool connector keeps the parallel data hierarchies (platform-side and tool-side) synchronous.
Ideally, a tool connector translates events produced by a third-party tool to the notification mechanisms of the meta model instance as well.
However, since many third-party tools to be connected to the platform do not provide an event notification mechanism via their public interfaces, this is not a requirement.
The tool connectors register the domain model instances resulting from the translation in the \emph{model registry} to make them available to other platform modules.
As best practice, tool connectors should provide domain model instances matching the domain model of the connected tool best possible.
These highly tool specific domain model instances ensure that as much data as possible is accessible from within the integration platform.

Working with tool specific domain models in higher level modules and apps is a rarely desirable thought.
A developer of a \emph{(client) platform app} (Figure~\ref{fig:abstract_architecture}) like the Timeline app as described above may want to work in the domain of events.
In turn, events are created based on domain data of integrated issue management systems or build servers.
To close the gap between these different models and their domains, the platform uses \emph{model mappers} (Figure~\ref{fig:abstract_architecture}) which map in terms of the Timeline app issues and build messages to events.

\emph{Platform apps} are the end-users of the provided models.
Like model mappers they take existing domain model instances as input, but contrary to them they do not produce other model instances as output.
Hence, platform apps can run in the same process as the integration platform (green painted platform apps), or in separate processes on other computers as client platform apps, see again Figure~\ref{fig:abstract_architecture}.
If they are run in separate processes at the client end, they can also run in separate instances of the platform,  which either contain a model registry.

\emph{Platform app connectors} (Figure~\ref{fig:abstract_architecture}) are used to connect several locally distributed instances.
Their task is to make the local model registry in form of a domain model instance containing further models available in the model registries of the remote end.
As a result, the business logic of the platform app can access remote models transparently in the same way as it would access local models.
Both sides of the connection, the client and the platform, can request the model registry of the other side, regardless of which side initiated the connection.
As as result, the platform has the possibility to access the data which is shared by the client.
This access can be restricted in implementations of the abstract architecture.

Platform app connectors can transfer arbitrary models that are based on the entity meta model.
In this process, all features of the meta model like the observability of changing data is preserved.
Contrary to tool connectors, platform app connectors are not tool specific but serialization and transmission technology specific.
That means, there can be a platform app connector that serializes meta model instances to binary data and transfers them directly over TCP/IP, or there can be a platform app connector that serializes to JSON and transfers via WebSockets.
So, platform app connectors are not specialized on special data types, for instance issue management data, as --by the abstraction provided by the entity meta model-- this becomes unnecessary.

\subsection{Entity Meta Model}\label{platformArchitectureMetaModel}
Data that should be shared between different modules is structured based on a special meta model called \emph{entity meta model}.
There are already many technologies used to structure data.
The entity meta model combines features of them to provide the foundation to structure data as needed in the integration platform.
For platform-independent description and exchange of information between systems, the Extensible Markup Language (XML) is already widely-used.
If XML was used as the meta model in the architecture, platform apps interested in data of other modules would have to parse XML documents.
This in turn means, the creator of a platform app would have to write technical code that transforms XML to some internal data model that can then be used to access this information more easily.
If data must be provided for other modules, code would have to be written to create XML documents.

The entity meta model instead defines interfaces used for accessing data.
Many programming languages have the concept of interfaces.
The entity meta model is therefore defined in a language independent way.
Contrary to XML, interfaces can usually be accessed directly by business code and therefore no technical transformation code is necessary within the app.

In many object oriented programming languages, classes are used to define domain models.
Like the entity meta model they are directly accessible from program code.
In instances respectively objects of classes the data is accessed via methods or fields.
Using classes as meta model for the platform without many additional restrictions was considered.
This way developers could define domain models in a familiar way, but further concepts are needed.

The meta model required for the platform must provide an abstraction level that enables the implementation of many generic functionalities.
Examples for these generic functionalities are caching that could be used to reduce the amount of access to potentially slow down third-party tools, polling, which can make data structures observable for changes, or transferring, which could be used to transfer data between the platform and external platform apps.
For the implementation of many of these generic functionalities, it is necessary to browse data of arbitrary concrete models that are based on the used meta model.
If for example a platform app connector should be able to transfer arbitrary instances of the meta model over the network, it would not be necessary to know the concrete method names that export the domain model data. Data browsing is hard to realize with classes.
The entity meta model has built-in support for browsing and accessing all data of an unknown instance of it.

In many apps based on the integration platform, it is not only required that domain data is readable and writable but it must also be possible to be informed when data changes.
Classes usually do not provide such notification mechanisms. Often event or message mechanisms are used to provide this feature, i.\,e., modules register to event queues or message topics to signalize the type of events they are interested in.
Senders write to these queues respective topics and the platform transmits the data to recipients.
This results in parallel data hierarchies: one for accessing data and one for event queues. Unfortunately this means, using this mechanism, the app developer would have to provide technical code that maps both hierarchies.
The entity meta model merges these hierarchies by providing a possibility to register observers for changes on most data knots in the domain model hierarchy.
This way the technical mapping code can be omitted and the app developer can focus on writing business code.

The notification mechanisms in the entity meta model are comparable to the JavaFX property mechanism.
But, for the platform the JavaFX mechanisms are, e.\,g. with their large inheritance hierarchy in the property classes, too complex to be usable as meta model.
Hence, we implement a simpler model; a developer of a generic functionality must be able to easily understand the whole meta model to consider all corner cases.

\begin{figure}[h!]
  \includegraphics[scale=0.63]{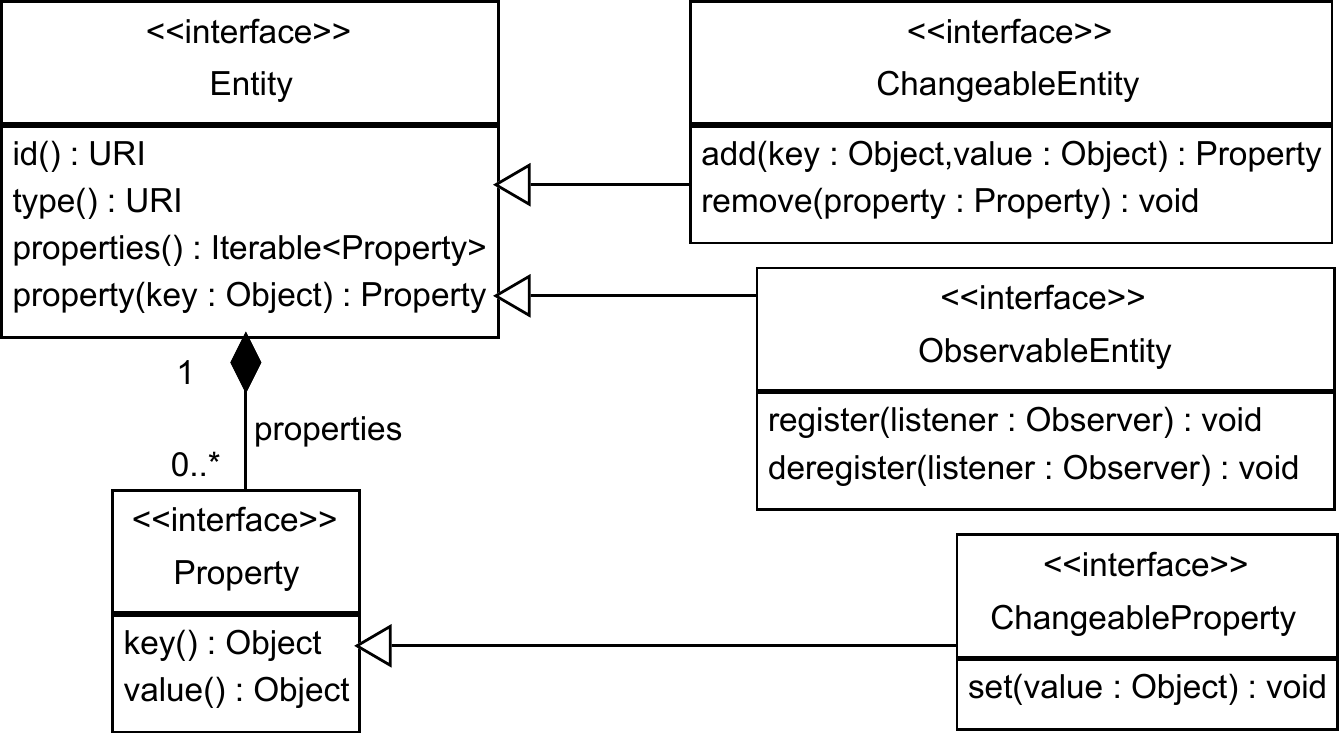}
  \centering
  \caption{An overview of the entity meta model. The core is the entity which has properties. Properties are referenced by a key or iterated over. Entities can be observable and/or changeable. The entity meta model is kept simple to be able to implement generic functionality like polling or caching.}
  \label{fig:entityMetaModel}
\end{figure}

Figure \ref{fig:entityMetaModel} gives a short overview of the elements of the meta model.
The entity meta model is designed with few concepts like interfaces and methods to be language agnostic.
Also the complexity is kept very low to be able to provide generic functionality easily. An \emph{entity} represents an aspect of a thing or subject which can be identified by an URI and the aspect that is described can be retrieved via the type method which also returns an URI.
E.\,g. for an issue management item there can be multiple entities in the platform that share the same id but have different types to describe it from different perspectives, e.g. as an issue or an event.
Entities consist of \emph{properties} and each property is identifiable by a key in the context of its entity and can hence be retrieved individually.
It is also possible to iterate over all properties of an entity.
The type of the properties values is restricted to basic types, i.\,e., string, integer, byte array or Entity.
An Entity is optionally observable for changing data and is optionally changeable from the outside.

%%%%%%%%%%%%%%%%%%%%%%%%%%%%%%%%%%%%%%%%%%%%%%%%%%%%%%%%%%%%%%%%%%%%%%%%%%%%%%%%

\section{DEVELOPING PLATFORM APPS}
\label{sec:DevelopingPlatformApps}

In this section it is discussed how the Timeline app as introduced above is implemented using the architecture discussed in the previous section to emphasize the simplicity and efficiency of our approach.
In the context of ALM tool integration, the Timeline app shows events of tools like Jira or Jenkins, which we explicitly consider to be the case here.
Further, we assume that connectors to these tools are already created by other developers and contributed to the platform, which is the typical practical scenario.
These connectors provide Jira respectively Jenkins specific models in the model registry.

As in most cases apps that are developed on top of the platform should work independent of a concrete tool, model mappers exist that map these tool specific models to generic domain models.
One mapper maps the Jira model to a change management model and another the Jenkins model to an automation model.
For the definition of generic domain models, existing vocabulary of OSLC models is reused wherever appropriate.
% http://open-services.net/wiki/ : Domain workgroups - vieleicht als Fußnote?
As described above, these mappings are bidirectional and also map change notifications.
When the Jira model notifies that a new issue was created, the change management model that is created through the mapping notifies that a new change request was created.
The created models are registered in the model registry to make them available to other modules of the platform.

In the development of the Timeline app, these generic domain models are taken as basis.
The Timeline app should show events that appeared in connected tools and therefore works in the domain of events.
In a first step, mappers are created that map from other domain models to the event domain model.
A mapping from the change management model to the event model ensures that notifications for new change requests are mapped to notifications for new events in the event model for example.
Besides the benefit of being able to work with a domain model that is appropriate for the app, the concept of model mapping ensures that later also events of other ALM domains can be displayed in the Timeline app by mapping their generic domain models to the event model.

The model mappings are executed on a server in the same integration platform instance that executes the tool connectors.
The Timeline app itself however is implemented as a JavaFX app that is run on the computer of the end user.
A generic platform connector that is running on both ends makes the event model of the server available in the model registry of the Timeline app client by communicating the changes via TCP/IP.
In JavaFX usually Java classes are used for the definition of the domain.
In the Timeline app they are also used for the definition of the event model.
Thereby, the same vocabulary is used that was used in the mappings.
Hence, the mapping between the entity meta model instance and the JavaFX instance of the event model can be automated with another generic functionality.

%%%%%%%%%%%%%%%%%%%%%%%%%%%%%%%%%%%%%%%%%%%%%%%%%%%%%%%%%%%%%%%%%%%%%%%%%%%%%%%%

\section{CONCLUSIONS}
\label{sec:Conclusions}

Within this work we identified important motivations for tool and data integration in the software development process. Besides the obvious advantages of more efficient development processes, a major further characteristics of our integration approach is the decoupling of integrated tools on one hand and applications running on the application platform on the other hand. This separation opens up endless possibilities as the replacement of vendor tools, while the every day processes and the used development tools, in our approach described as platform apps, do not change. Also, the app concept and the functionality store to provide those apps helps to use what the community provides for certain development processes or to also replace apps without changing or replacing the applied tools. 

Moreover, in case no app is available, the functionalities are accessible directly by the programmer to develop new functionalities -- and apps -- which in turn can be published in the function store. Hence, the platform opens new possibilities and more flexibility in tool usage and integration and opportunities for creating new tool features using existing ones. Given this rich environment, coupling arbitrary development tools with the platform and mapping their data, functionalities and events to internal representations and making them accessible by a large number of high-level platform apps, gives rise to completely customized development scenarios and development processes. If the available characteristics and functionalities are applied right, the development team works exactly with the tools (3rd party tools and apps) needed for a specific development project and most probably with different tools in different projects. This is agile development at its best.

We introduced examples for possible platform apps as an Information Dashboard or a Project Cockpit, which make the idea transparent. We outlined the requirements on the platform and propose an architecture. The entity meta model has been described as a solution for defining own logic data models used by platform apps. According to the platform architecture and the meta model we described how platform apps are build and how the platform can be used for the implementation of a platform app based on the vanilla case of the Timeline App, which has been implemented. The current prototype implements the concepts that are described within the paper and can be demonstrated. 

\section{FUTURE WORK}
\label{sec:FutureWork}
 
For further platform apps, as e.\,g. described in the motivation section, the platform has to be extended. As specified, there are many systems or tools left to be integrated. For many systems, connectors have to be implemented. Writing a generic OSLC connector would be a clever way to save efforts realizing this, since this would support all tools which support OSLC.

The sharing of functionalities can be seen as a generalization of the successful app concept. For sharing created functionality, the functionality store has to be further developed and started on the web. The store and the platform need intelligent search functionalities, too, as described above. We described that the model can be enriched with semantic information.

Another very important step is the decoupling of the UI from concrete client implementations. At the moment a platform app would be implemented in a client technology like JavaFX using the platform API. That is to say the UI is developed by the means of JavaFX. Actually, UI components should be coupled with the functionality and special data models. That is a client technology neutral representation is needed, which leads us to the wish of having predefined generic and specialized view components, especially editors. A generic model data browsing component would be able to show the structure and general information of models, whereas specialized viewers or editors work only on special model (entity) types.

There are many more possible developments to mention. A very last one to be mentioned here is again about tool support, but now in the context of platform app development. Developers need tools supporting the development of platform apps. This may be an editor component, preferably as a web application, providing auto-completion and syntax support that is connected to the functionality repository of the platform.

\newcommand{\noopsort}[1]{} \newcommand{\printfirst}[2]{#1}
  \newcommand{\singleletter}[1]{#1} \newcommand{\switchargs}[2]{#2#1}

\end{document}